\documentclass[]{spie}  

\usepackage{amsmath,amsfonts,amssymb}
\usepackage{graphicx}
\usepackage[colorlinks=true, allcolors=blue]{hyperref}

\title{The Simons Observatory: Development and Validation of the Large Aperture Telescope Receiver}

\author[1]{Tanay Bhandarkar}
\author[2]{Sanah Bhimani}
\author[3,4]{Gabriele Coppi}
\author[1]{Simon Dicker}
\author[1]{Saianeesh K.	Haridas}
\author[5]{Kathleen Harrington}
\author[1]{Jeffrey Iuliano}
\author[6]{Bradley Johnson}
\author[1]{Anna M. Kofman}
\author[7]{Jack Lashner}
\author[8]{Jenna Moore}
\author[2]{David V. Nguyen}
\author[1]{John Orlowski-Scherer}
\author[1]{Karen Perez Sarmiento}
\author[1]{Julia Robe}
\author[9]{Maximiliano Silva-Feaver}
\author[10,1]{Robert J. Thornton}
\author[11]{Yuhan Wang}
\author[12]{Zhilei Xu}

\affil[1]{Department of Physics and Astronomy, University of Pennsylvania, Philadelphia, PA, USA}
\affil[2]{Department of Physics, Yale University, New Haven, CT, USA}
\affil[3]{Department of Physics, University of Milano-Bicocca, Milano, Italy}
\affil[4]{National Institute for Nuclear Physics , Sezione di Milano-Bicocca, Milano, Italy}
\affil[5]{Department of Astronomy and Astrophysics, University of Chicago, Chicago, IL, USA}
\affil[6]{Department of Astronomy, University of Virginia, Charlottesville, VA, USA}
\affil[7]{Department of Physics and Astronomy, University of Southern California, LA, CA, USA}
\affil[8]{School of Earth and Space Exploration, Arizona State University, Tempe, AZ, USA}
\affil[9]{Department of Physics, University of California San Diego, La Jolla, CA, USA}
\affil[10]{Department of Physics, West Chester University of Pennsylvania, West Chester, PA, USA}
\affil[11]{Department of Physics, Princeton University, Princeton, NJ, USA}
\affil[12]{MIT Kavli Institute, Massachusetts Institute of Technology, Cambridge, MA, USA}

\authorinfo{Corresponding author: Tanay Bhandarkar\\ E-mail: tanayb@sas.upenn.edu}

\begin{document} 
\maketitle

\begin{abstract}
The Simons Observatory (SO) is a ground-based cosmic microwave background (CMB) survey experiment that consists of three 0.5\,m small-aperture telescopes (SATs) and one 6\,m large-aperture telescope (LAT), sited at an elevation of 5200\,m in the Atacama Desert in Chile. In order to meet the sensitivity requirements set for next-generation CMB telescopes, the LAT will deploy 30,000 transition edge sensor (TES) detectors at 100\,mK across 7 optics tubes (OT), all within the Large Aperture Telescope Receiver (LATR). Additionally, the LATR has the capability to expand to 62,000 TES across 13 OTs. The LAT will be capable of making arcminute-resolution observations of the CMB, with detector bands centered at 30, 40, 90, 150, 230,
and 280\,GHz. We have rigorously tested the LATR systems prior to deployment in order to fully characterize the instrument and show that it can achieve the desired sensitivity levels. We show that the LATR meets cryogenic and mechanical requirements, and maintains acceptably low baseline readout noise. 

\end{abstract}


\section{Introduction}
Observations of the cosmic microwave background (CMB) over the last several decades have been instrumental to the development of modern cosmological models, such as $\Lambda$-CDM. Our understanding of the CMB has evolved through full-sky surveys conducted by space-based satellite experiments such as COBE, WMAP\cite{bennett2013} and Planck\cite{planck2020}. Large aperture ground based experiments, such as the Atacama Cosmology Telescope (ACT)\cite{thornton2016} and the South Pole Telescope (SPT)\cite{Ruhl2004}, have improved our ability to make small angular measurements of the CMB. These small angular measurements have offered insights into the development of large scale structure, the upper bound on the sum of neutrino masses, the Sunyaev-Zeldovich effects (SZ) and primordial power spectra. Other ground based telescopes, such as CLASS\cite{xu/etal:2020c} and BICEP/Keck, utilize small apertures to primarily measure the B-mode polarization signal. \par
\begin{figure}
    \centering
    \includegraphics[scale=0.8]{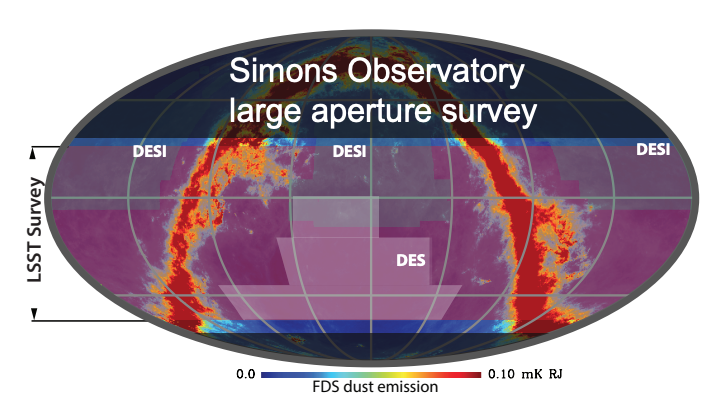}
    \caption{Planned sky coverage for the LAT \cite{SO_Goals}.}
    \label{fig:LAT_sky_coverage}
\end{figure}

The Simons Observatory (SO) is a next-generation, ground-based CMB experiment, consisting of three 0.5\,m small-aperture telescopes (SATs) and one 6\,m large-aperture telescope (LAT). These telescopes will be deployed at an elevation of 5200\,m in the Atacama Desert of Chile. The LAT will utilize a crossed-Dragone optical design, enabling a large focal plane for the telescope. The Large Aperture Telescope Receiver (LATR) will occupy the LAT's focal plane with $\sim$30,000\, transition edge sensor (TES) bolometers across 7 optics tubes (OT), and has the capability to upgrade to $\sim$62,000 TES across 13 OTs. Each OT is specifically designed for a separate frequency band (with targeted band centers); low-frequency (30 and 40\,GHz), medium-frequency (90 and 150\,GHz) and ultra-high frequency (230 and 280\,GHz). Over the course of it's observing campaign, the LAT will scan 40\% of the sky at arcminute resolution, with significant overlap with other surveys at complementary wavelengths, such as the Dark Energy Survey (DES)\cite{DES}, the Dark Energy Spectroscopic Instrument (DESI)\cite{DESI2016} and the Legacy Survey of Space and Time (LSST) with the Vera Rubin Observatory\cite{LSST__2019} (Figure \ref{fig:LAT_sky_coverage}). The field of view of the LATR will be approximately $7.05^\circ$ in diameter. The LAT will focus on small-angular sciences, including measurements of the temperature, polarization and lensing power spectra, the detection of galaxy clusters through the thermal SZ effect, the detection of extra-galactic point sources and transient event monitoring \cite{SO_Goals}. \par

\section{LATR Design}
\begin{figure}[h]
    \centering
    \includegraphics[scale=0.6]{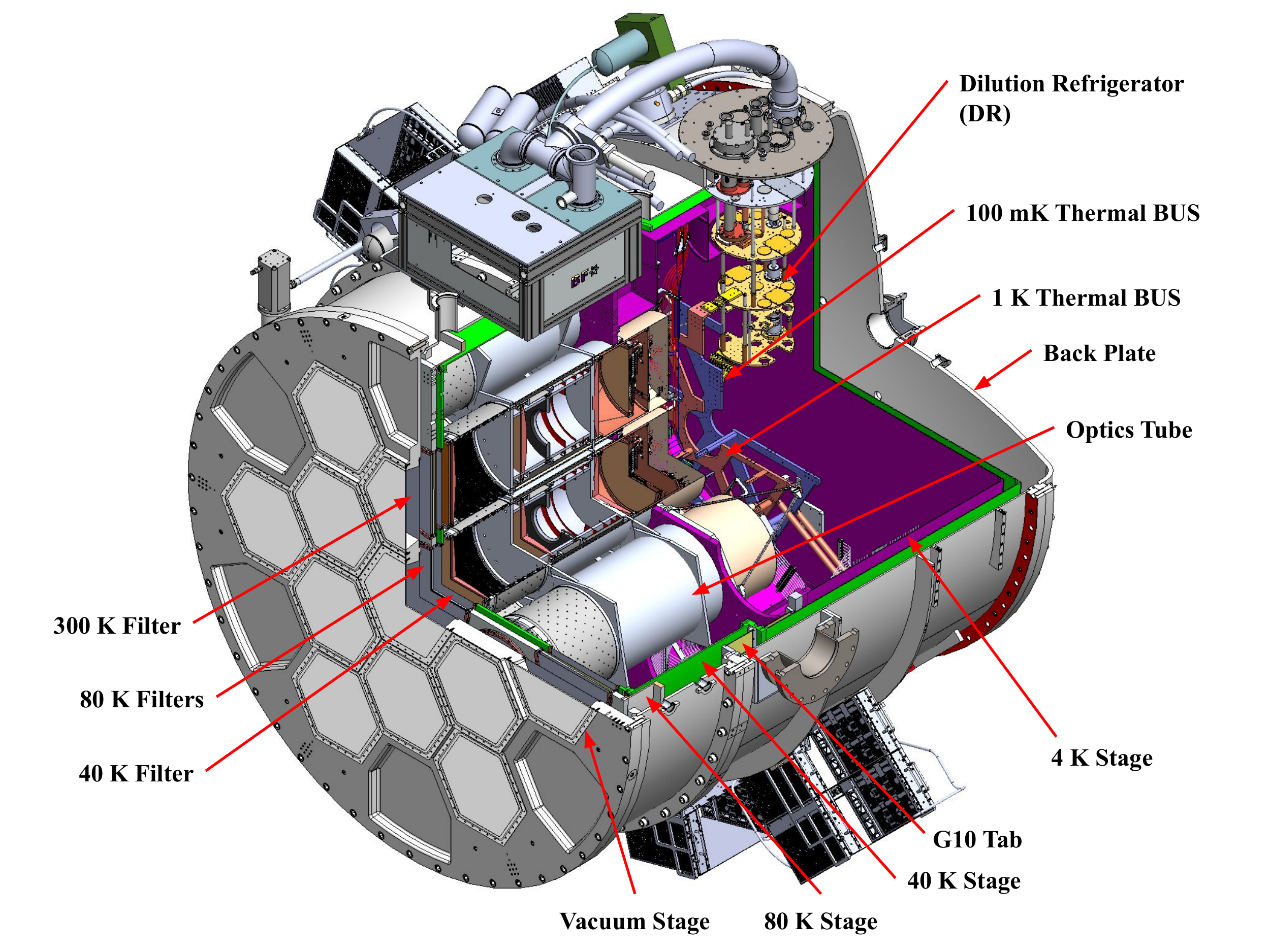}
    \caption{A cross-sectional view of the LATR, with the 300\,K (gray), 80\,K (gray), 40\,K (green), 4\,K (pink) stages all visible. Additionally, the 1\,K BUS (brown) and 100\,mK BUS (indigo) can be seen at the back of the LATR, close to the DR. This view also displays the entire filter stack in an optics tube. On the exterior of the LATR, the SMuRF readout crates can be seen mounted at different locations, as well as the DR's turbo pump station on top. \cite{Zhu_2021}.}
    \label{fig:latr_xsection}
\end{figure}

The LATR consists of 5 separate nominal temperature stages; 300\,K, 80\,K, 40\,K, 4\,K, 1\,K and 100\,mK. 
We will describe the design of the cryostat in this section, and further details can be found in other publications.\cite{Zhu_2021, Xu_2021}. Figure \ref{fig:latr_xsection} shows the cross-sectional view of a full LATR. 
\subsection{Cryo-mechanical and Optical Design}
\begin{figure}[h]
    \centering
    \includegraphics[scale=0.65]{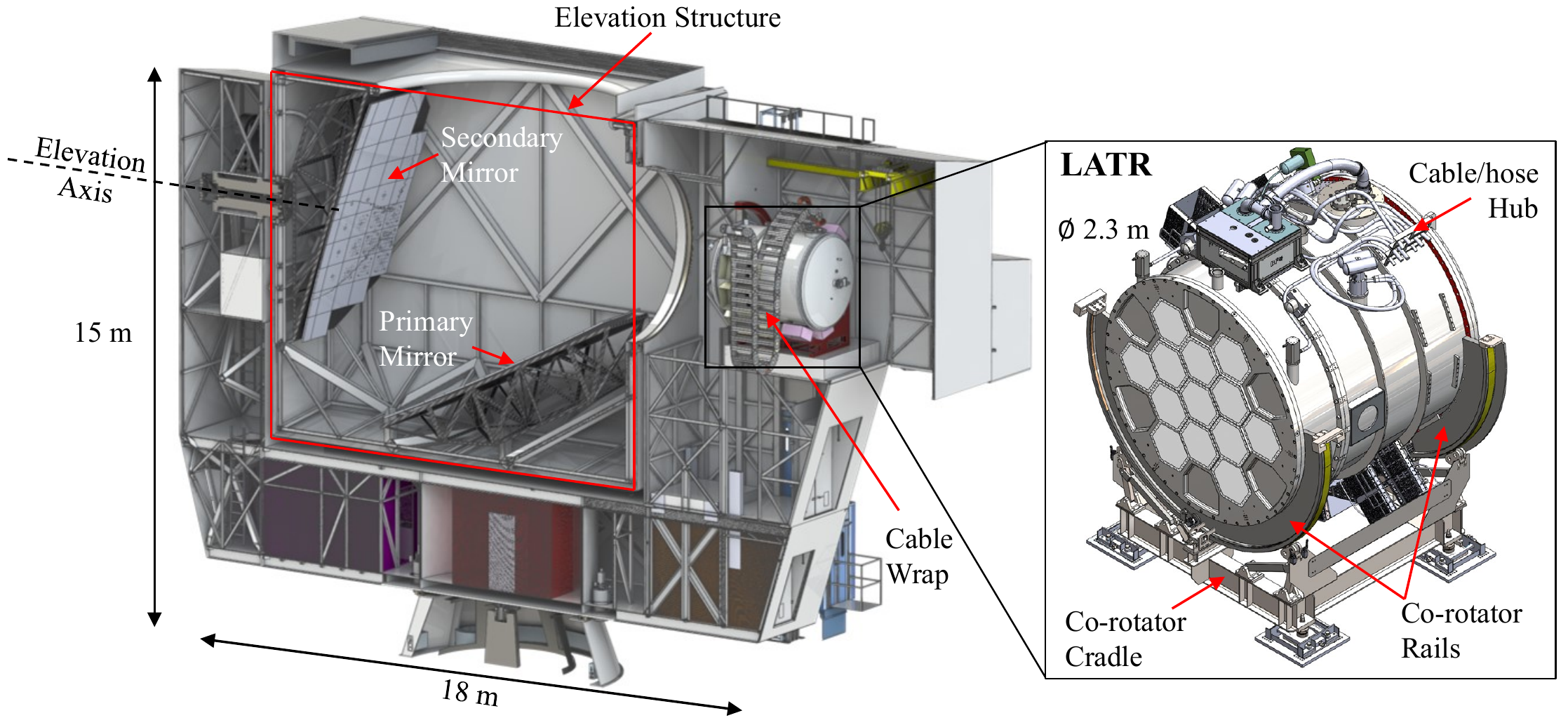}
    \caption{Computer aided design models of the LAT (left) and LATR (right). The LATR will sit on the co-rotator rails inside the LAT's receiver cabin, enabling it to rotate along with the LAT's elevation structure. The LAT sits on a large conic base that will rotate the telescope azimuthally. Figure from Xu et al. 2021.\cite{Xu_2021}}
    \label{fig:LAT_Plus_LATR}
\end{figure}

The 300\,K vacuum shell is 2.4\,m in diameter, and 2.6\,m in length, made of aluminum-6061. The front plate is 6\,cm thick with a honeycomb pattern for 13 anti-reflective (AR) coated ultra-high-molecular-weight polyethylene (UHMWPE) 1/8" thick windows. The thicknesses of the vacuum shell, front and back plates, and windows were determined following a finite element analysis (FEA) study into the structural requirements at atmospheric pressures \cite{orlowski_2018}. Behind each UHMWPE window sits a double-sided IR blocker (DSIR), which reflects thermal radiation away from the the colder stages \cite{ade_2006}. The 300\,K stage will mate the LATR to the LAT through a pair of co-rotating rails, that turn the receiver as the elevation structure of the telescope rotates (see Figure \ref{fig:LAT_Plus_LATR}).  \par

The 80\,K stage consists of the 80\,K filter plate and the 80\,K shell. This filter plate holds a DSIR and an anti-reflection (AR) coated alumina filter\cite{Golec_2020} per OT. This set of filters absorb a significant fraction of radiative power that comes in through the 300\,K windows. The 40\,K stage holds a single 40\,K DSIR per OT.
The 4\,K stage consists of a back lid, and filter plate and a radiation shield. Because the OTs mount on this stage, the 4\,K filter plate is significantly thicker than the other cold filter plates to better support the weight of the OTs and improve the thermal path from the pulse tubes to the OTs. Additionally, the 80\,K, 40\,K and 4\,K stages are all wrapped in 30 layers of multi-layer insulation (MLI), preventing thermal radiation from reaching the colder stages.  \par
The 80\,K stage is cooled by two PT-90 pulse tube coolers, while the 40\,K and 4\,K stages are cooled by a pair of two-stage PT-420 pulse tube coolers\footnotemark{}\footnotetext{https://www.cryomech.com/}. All pulse tube heads are coupled to their respective cold stage by a set of copper straps. Each copper strap is a pair of oxygen-free, high conductivity (OFHC) copper plates, bridged by copper braids.  \par
The requirements for the LATR design demands the  ability to balance mechanical integrity with thermal isolation between the different temperature stages.  To accomplish this, the LATR utilizes three separate rings of G-10 tabs. These tabs mechanically connect the 300\,K stage to the 80\,K stage, the 300\,K stage to the 40\,K stage, and the 40\,K stage to the 4\,K stage. Each G-10 tab consists of two aluminum "feet" epoxyed to a sheet of G-10. The feet are then bolted to their respective cold stage cold stages.\par
\begin{figure}[ht]
    \centering
    \includegraphics[scale=0.3]{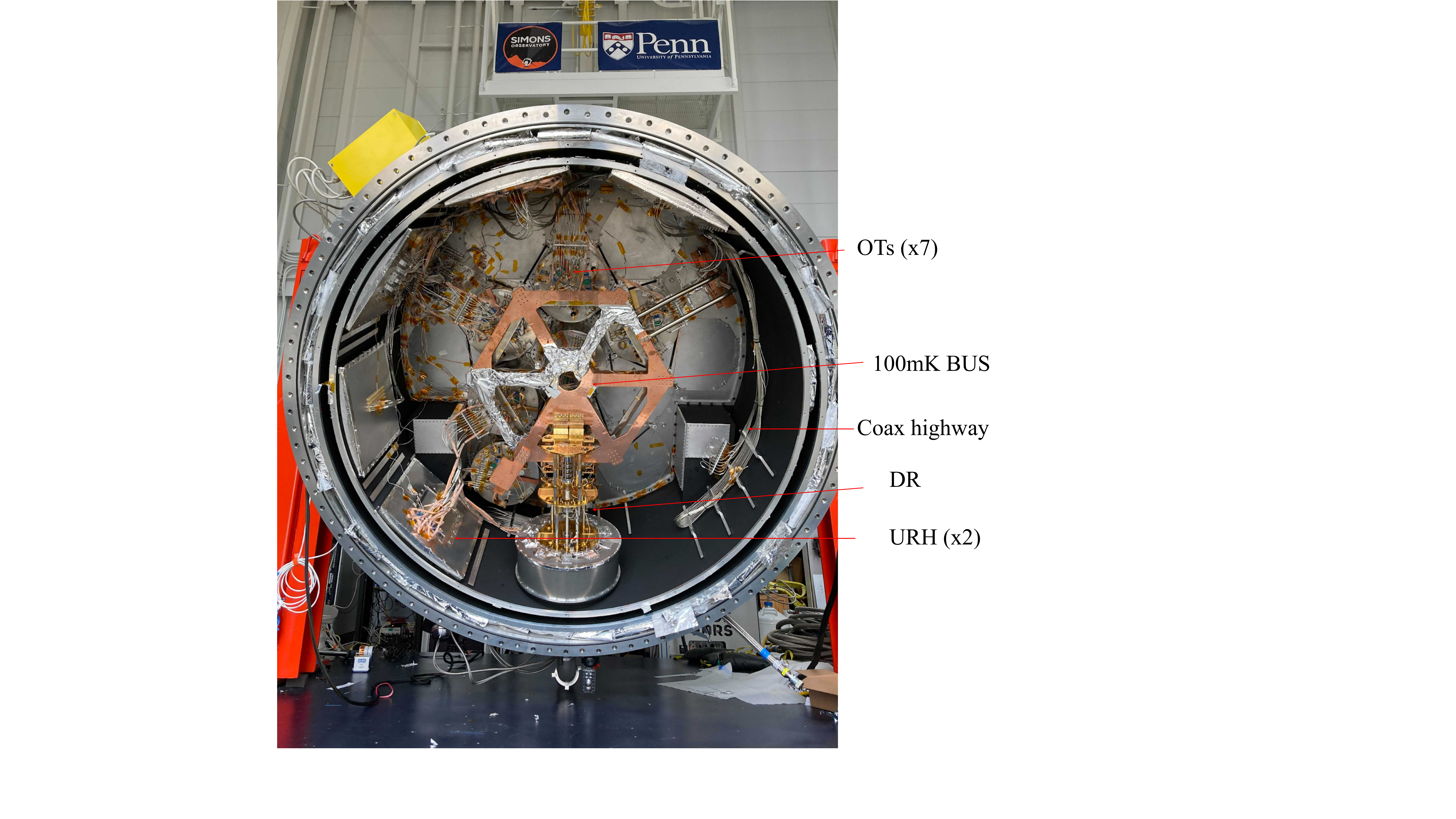}
    \caption{The LATR with its back plates removed. seven optics tubes (OTs) are installed in this photo. Each OT is connected to one of the two universal readout harnesses (URH) via a lengthy coaxial connection, colloquially named the "Coax highway" at 4\,K.}
    \label{fig:open_latr}
\end{figure}

The TES detectors are cooled to 100\,mK in order to reduce thermal noise and improve sensitivity. To avoid larger thermal gradients across OT 100\,mK stages, the LATR implements a wheel-like structure for its 1\,K and 100\,mK thermal back-up structure (BUS) cold stage (see Figure \ref{fig:open_latr}). The thermal BUS stages are both made of OFHC copper. Thermal FEA predicts a gradient of $\sim$ 5\,mK across the entire 100\,mK BUS. These cold stages are cooled with a Bluefors\footnotemark{}\footnotetext{https://bluefors.com/} model LD 400 dilution refrigerator (DR). The thermal BUS stages are coupled to their respective stages of the DR through gold-plated, braided copper straps. In order to cool the 1\,K stage and focal plane baseplates (FPB) that supports the detector arrays in each OT, the LATR utilizes a pair copper rod cold fingers (one on the FPB(100\,mK) or radiation lid (1\,K), one on each stage of the thermal BUS), bridged by a separate  gold-plated, braided copper strap. The 1\,K BUS is mounted to the 4\,K plate by six twill-ply carbon fiber tube tripods. The 100\,mK BUS is supported by a truss of carbon fiber connecting to the 1\,K BUS.  
\subsection{Readout and Detector Design}
In order to achieve the sensitivity goals of SO, the LATR will field $\sim$62,000 TES detectors in its final configuration. To efficiently read out this large volume of sensors, the LATR implements a microwave squid multiplexing readout system, carrying tones along coaxial lines from the 300\,K stage down to the the 100\,mK universal focalplane modules (UFMs), and back out. The expectation is to be able to read out $\mathcal{O}(10^3)$ detectors per coaxial line.\cite{sathyanarayana2020} UFM design has been detailed in prior publications.\cite{McCarrick_2021}
\subsubsection{Cold Readout design}
The cold readout chain carries RF (4-6\,GHz) tones and DC ($<$1\,MHz) biases output by warm readout electronics, into the cryogenic receiver, to the cryogenic detectors and back out to the SMuRF. From 300\,K to 4\,K, the LATR readout system is contained within the Universal Readout Harness (URH), an assembly that is used throughout SO receivers\cite{moore2022}. In a single URH, there are up to 12 bias channels and 24 coaxial channel. A bias channel is a 50 pin input that carries 2 flux ramp line pairs, 12 detector bias pair lines, and 4 cryogenic amplifier bias lines. Each coaxial channel consists of an input connection and output connection, and the respective cold readout components. Ultimately, each UFM is read out through two coaxial chains, and each bias channel in the URH is coupled to two separate coaxial chains. In sum, a single, fully populated LATR URH can read out up to 12 UFMs. \par 
At the 4\,K stage, a series of coaxial connections (referred to as coax highways from here out) and DC bias lines connect the URHs to individual OTs. The input lines in the OTs consist of a series of attenuators to optimize tone power that reaches the resonators, and DC blocks that provide thermal isolation in coaxial connections between cold stages. On the output line from the UFMS, there are low-noise amplifiers (LNA) to amplify the output signal, and isolators to prevent standing waves between the UFM and the LNAs. In parallel to the coaxial connections, each readout chain contains DC bias lines that travel all the way down to the UFM.
These bias lines  provide voltage biases for the LNAs, detector biases and a ramp signal for the flux ramp signal used by warm electronics to linearize the resonator frequency shifts produced by the detectors.\cite{mates2012}
\subsubsection{Warm Readout design}
Each TES on an UFM is coupled to a resonator at a unique resonant frequency, between 4-6\,GHz. SO will uses the SLAC Microresonator Radio Frequency (SMuRF) electronics to generate these 4-6\,GHz frequency tones, transmitted through the cryogenic RF components via coaxial lines. Up to 147 TES detectors share the same voltage bias line and 882 detector channels share the same resonator flux ramping lines, both of which are generated by the SMuRF electronics. \cite{henderson2018}\par
4-6\,GHz RF tones are generated in two Advanced Mezzanine Cards (AMC) seated in a carrier card. Every AMC is connected to a single URH coaxial channel, along with a warm amplifier mounted on the URH. Each carrier card is coupled to a Rear Transition Module (RTM), which is responsible for generating all the low frequency signals (detector biases, flux ramps, amplifier biases). The RTMs are connected to a single cryostat card (cryocard), which is also mounted above the URH. The cryocards convert the TES bias and flux ramp signals from the RTM's low impedance voltage sources to high impedance current sources. It also provides regulated power supplies, bias line filtering and gate bias voltages to LNAs. \par
In order to reduce the effects of ambient RF noise and minimize cable loss, the SMuRF crates are mounted directly on the LATR shell. There are extensive ongoing studies into effects of thermal and vibrational environments on the warm readout system. 

\section{Validation}
\subsection{Cryogenic Validation}
In order to meet detector sensitivity and noise requirements, it is crucial for the LATR to cool every OT FPB to a stable 100\,mK level. Our 100\,mK loading estimates for various configurations are derived from baseline load curves. In other words, we first characterized our DR by applying a range of thermal loads to its mixing chamber and still line stages. We then methodically built up the LATR by adding different cryogenic components; the thermal BUS, dark OTs, and finally optically coupled OTs. Past studies of the LATR have calculated the thermal load of the 100\,mK thermal BUS to be $\sim$20$\mu$W, and the thermal load of a single dark OT to be$\lesssim$5$\mu$W, and the temperature of the OT's FPB varied by less than 0.3\,mK over 24 hours \cite{Zhu_2021} (In this study, "dark" indicates no cold optics below 40\,K, and an aluminum blank over the OT's 4\,K filter cells). Here we will detail the LATR cryogenic thermal loading values with 7 dark OTs, and then with 1 optical. Details on 80\,K, 40\,K and 4\,K loading can be found in Tables \ref{table:80K Loading},\ref{table:40K Loading}  and \ref{table:4K Loading}. \par
\subsubsection{7 Dark OTs} \label{section:7_dark_ots}
\begin{figure}[h]
    \centering
    \includegraphics[scale=0.5]{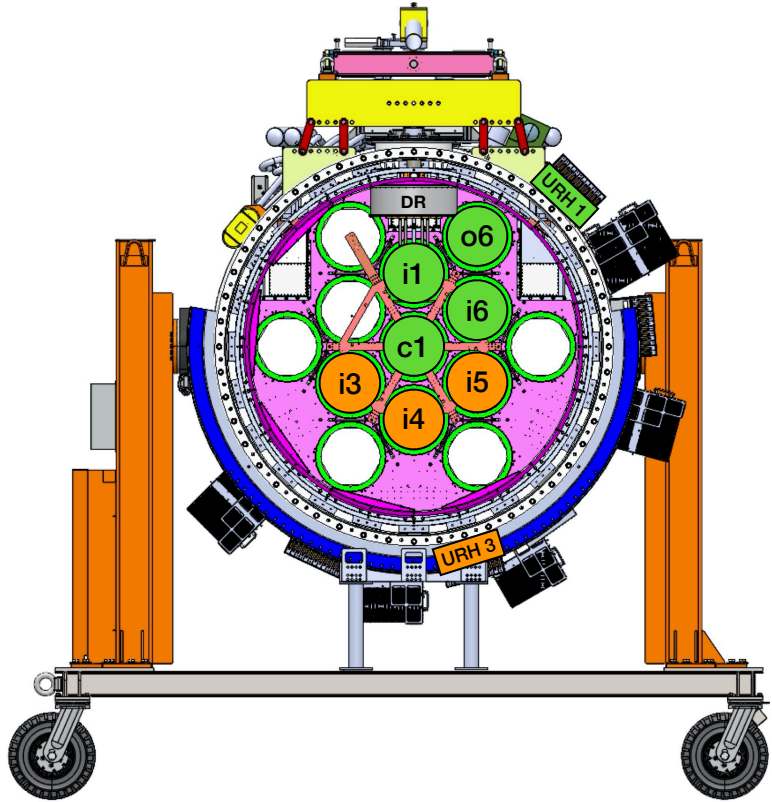}
    \caption{The deployment configuration of the LATR with 7 optics tubes (OTs), as viewed from the rear of the receiver. The color of the OT indicates which universal readout harness (URH) it will be read out through. OTs i1, i3, i4 and i6 will be MF tubes. OT c1 and i5 will be UHF tubes. OTo6 will be the only LF tube intially deployed.}
    \label{fig:7OT_distribution}
\end{figure}

A significant milestone in LATR lab testing was the cryo-mechanical integration of 7 dark OTs, in the deployment configuration positions, as shown in Figure \ref{fig:7OT_distribution}. In this configuration, the OTs were covered with an aluminum plate at the 4\,K filter stack. At the 100\,mK stage, the DR mixing chamber plate stabilized at about 43.9\,mK. Based on prior load characterizations, we estimated that each OT on average exerts $\sim$7.1$\mu$W at 100\,mK through parasitic loading. The hottest FPB in this configuration was  $\sim$75\,mK. Projecting these numbers out to a 13 OT configuration, we estimate that the hottest FPB would be $\leq$ 105\,mK. At this temperature, we run the risk of putting detectors over their expected baseline temperature. In order to rectify this, we plan to install additional thermal straps coupling the DR to the 100\,mK thermal BUS, thus improving the transfer of cooling power to the BUS. \par
 At the 40\,K and 4\,K stage, we estimate $50.9\pm0.9$W and $1.27\pm0.06$W respectively. Extrapolating this loading to 13 OTs, we find that the 40\,K stage and 4\,K stage will cool to $\sim 32.5$K and $3.95$\,K respectively. Both of these values fall within specifications.  At 80\,K, we measure $87.9$W. For 13 OTs, these loading values suggest at filter plate temperature of $\sim$82\,K, which we expect will be sufficiently cold. \par
\begin{figure}[h]
    \centering
    \includegraphics[scale=0.75]{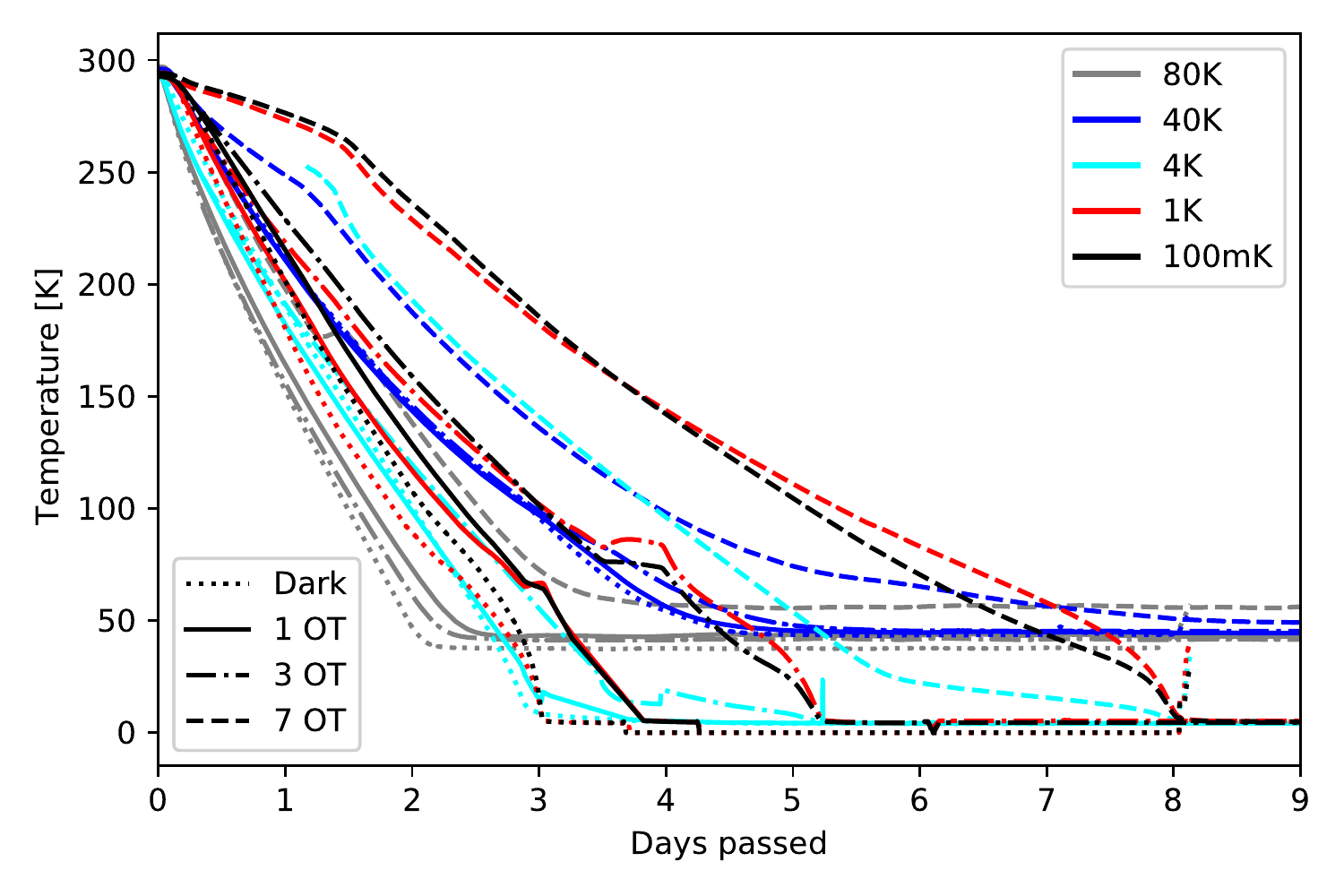}
    \caption{Cooldown times for the LATR in various configurations. In all of these configurations, the optics tubes (OT) are capped at 4\,K.} 
    \label{fig:full_comparison}
\end{figure}

 When the LATR is finally in the field, cooldown speed is of great importance; the more time it takes to reach base temperature, the less time is spent observing. With 7 OTs, we found that it takes approximately 8 days for all of the LATR cold stages to reach base temperature (Figure \ref{fig:full_comparison} ). The cooldown time is dominated by the drop to 4\,K on the coldest stages, while the 80\,K and 40\,K stages reach their respective base temperatures within 5 days. We utilize a mechanical heat switch assembly, which creates thermal shorts between the 4\,K plate and the 1\,K and 100\,mK BUS. Without this assembly, simulations suggest it would the LATR to take over 35 days to reach base temperature. \cite{coppi_2018}. We expect that with 13 OTs, the heat switches will cut that time in half. \par
\subsubsection{1 Optical OT}
\begin{figure}
    \centering
    \includegraphics[scale=0.5]{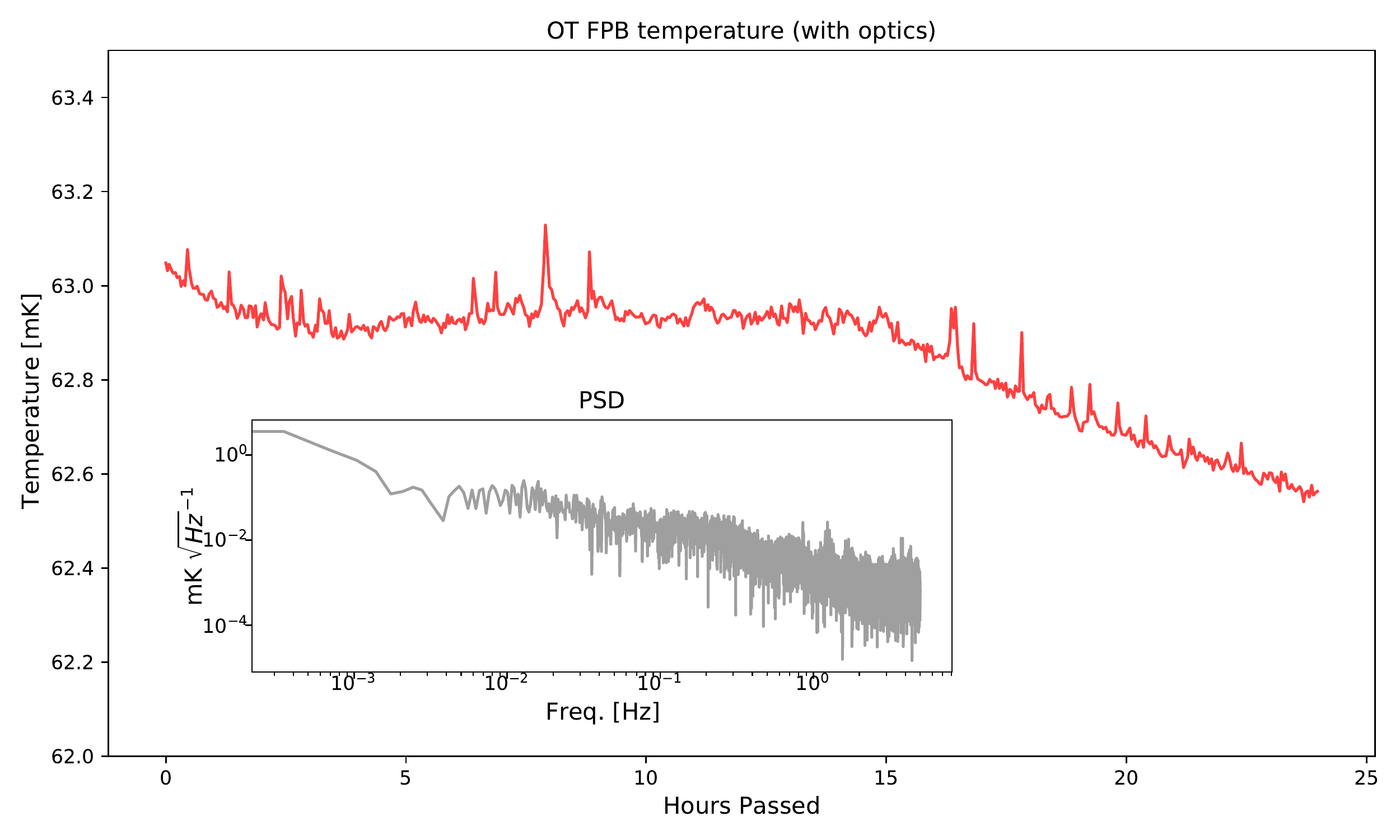}
    \caption{Temperature of the OT focal plane baseplate (FPB) behind the full optical stack, over the course of 24 hours. During this time, the cryostat was left untouched. The inset plot displays the power spectral density (PSD) of that thermometer during the same period of time. The flatness of the PSD indicates the stability of the temperature.}
    \label{fig:fpb_stability}
\end{figure}

After completing dark cryogenic testing of the OTs, we installed one OT equipped with the full cold filter and lens stack at the OTi6 position (Figure \ref{fig:7OT_distribution}). This test configuration included the full complement of readout hardware, but did not include full detector array assemblies. In lieu of detector arrays, we installed three copper mock arrays with Al feedhorns, to mimic the expected mass and collecting area of an UFM at that stage. The base temperature of the OT's FPB was $\sim$55\,mK, with very little variation over the course of 24 hours (see Figure \ref{fig:fpb_stability}). Based on thermal load curves taken when the LATR was dark, and only cooling the 100\,mK thermal BUS, we determined that the thermal load of a single optical OT $\leq$6.0\,$\mu$W. Thermal simulations predict that the full detector arrays will add $\sim$1.0\,$\mu$W per OT when they are voltage biased. This loading number is higher than predicted at the 100\,mK stage; we expect additional straps mentioned in section \ref{section:7_dark_ots} to augment cooling transfer sufficiently. \par
Based on our cryogenic testing, there appear to be small discrepancies between loading numbers extracted from different cooldowns. There are a couple explanations for this phenomenon. One, comparing thermal loading between cooldowns assumes that cold strapping is assembled identically between tests. In reality, this is difficult to ensure; the strapping surfaces experience wear and bolts may not always be tightened to the same torque. Second, it is difficult to precisely measure the thermal loading to sub-$\mu$W levels due to variations in hardware and noise in thermometry data. \par
\begin{table}
\begin{center}
   \begin{tabular}{c|c|c}
    Test Configuration & 80\,K Plate Temperature & Thermal Loading    \\
    \hline
    \hline\
    Dark & $35.4 - 38.8$\,K & $22_{-1}^{+1}$\,W \\
    2 Filter Stacks & $37.2-43.1$\,K & $35_{-1}^{+1}$\,W \\
    3 Filter Stacks & $41.0-48.7$\,K & $42_{-1}^{+1}$\,W \\
    7 Filter Stacks & $46.8-58.3$\,K & $69_{-1}^{+1}$\,W \\
    \end{tabular}
    \caption{\label{table:80K Loading}Loading estimates on the 80\,K stage, based on temperatures and load curves taken from the two PT-90 cryocoolers. The 80\,K Plate temperatures represent the range of temperatures between the warmest point on the filter point to the coldest pulse tube.}
\end{center}
\end{table} 
\begin{table}[h]
\begin{center}
   \begin{tabular}{c||c|c}
    Test Configuration & 40\,K Plate Temperature & 40\,K Thermal Loading  \\
    \hline
    \hline\
    Dark* & $29.3-26.8$\,K & $33_{-1}^{+1}$\,W \\
    2 Filter Stacks &$27.2-48.5$\,K & $34_{-1}^{+1}$\,W \\
    3 Filter Stacks & $27.3-49.1$\,K  & $36_{-1}^{+1}$\,W \\
    7 Dark OTs &$30.7-53.7$\,K & $51_{-1}^{+1}$\,W\\
    1 Optical OT & $27.6-49.1$\,K & $37_{-1}^{+1}$\,W \\
    \end{tabular}
    \caption{\label{table:40K Loading}Loading estimates on the 40\,K, based on temperatures and load curves taken from the first stages of the two PT-420 cryocoolers. The asterisk indicates a cooldown with no URH installed. Test configurations with OTs assume at least 7 full filter stacks down through 40\,K. Plate temperatures represent the range of temperatures between the warmest point on the filter point to the coldest pulse tube. }
\end{center}
\end{table} 
\begin{table}[h]
\begin{center}
   \begin{tabular}{c||c|c}
    Test Configuration & 4\,K Plate Temperature & 4\,K Thermal Loading    \\
    \hline
    \hline\
    Dark*  & $2.74-5.02$\,K &$0.8_{-0.1}^{+0.1}$\,W \\
    2 Filter Stacks &$3.04-6.68$\,K & $0.8_{-0.1}^{+0.2}$\,W\\
    3 Filter Stacks & $2.90-5.15$\,K & $1.3_{-0.2}^{+0.2}$\,W \\
    7 Dark OTs  &$3.09-6.44$\,K & $1.27_{-0.1}^{+0.1}$\,W\\
    
    1 Optical OT  & $2.83-5.15$\,K&\ $0.8_{-0.1}^{+0.1}$\,W\\
    \end{tabular}
    \caption{\label{table:4K Loading}Loading estimates on the 4\,K stages, based on temperatures and load curves taken from the second stage of two PT-420 cryocoolers. The asterisk indicates a cooldown with no URH installed. Test configurations with OTs assume at least 7 full filter stacks down through 40\,K. Plate temperatures represent the range of temperatures between the warmest point on the filter point to the coldest pulse tube. }
\end{center}
\end{table} 

\subsection{Readout Validation}
Given the complexity of the LATR's readout design, it is important to show that we are able to maintain acceptably low readout noise when deploying UFMs. Past studies have shown TES white noise levels are close to expected models in a single pixel box, containing 6 TES and 65 resonators\cite{Xu_2021}. LATR MF UFM in situ testing is planned before deployment. This testing includes establishing baseline receiver noise levels, and studies into the effects of various environmental factors, such as vibrations, cryogenic thermal instability, ambient RF backgrounds and warm readout thermal coupling. Optical testing, including holography, cross-polarization measurements and beam mapping, has been conducted in a testbed cryostat \cite{harrington_2020}. 

\section{Conclusion}
The Simons Observatory (SO) Large Aperture Telescope (LAT) will be a sensitive millimeter-wave telescope with a large focal plane. The Large Aperture Telescope Receiver (LATR) will occupy this focal plane with up to $\sim$62,000 transition edge sensor (TES) detectors, across 13 modular optics tubes (OT). Each OT will contain 3 universal focal plane module (UFM) arrays, and the respective readout and cold optics colder than 4\,K. Due to the sensitivity requirements, the LATR must balance a complex optical and readout design with a robust cryo-mechanical design. Through several rounds of cryogenic testing, we show that the LATR will be capable of cooling down cold optical elements, readout assemblies and detector arrays for 7 OTs for its initial deployment. For 13 OTs, we will add additional cold strapping to ensure we are comfortably within cryogenic specifications. In order to optimize TES detector noise levels, the LATR must achieve a stable 100\,mK base temperature. Cryogenic lab testing of the receiver shows that it is capable of cooling the focal plane baseplate of a single optically coupled optics tube (OT) to a base temperature of $\sim$54\,mK. Testing of multiple optically coupled OTs with fully functional UFMs is underway. \par
Additionally, lab testing has demonstrated the viability of the LATR's readout system, including its ability to track resonators and bias detectors. Further laboratory readout testing is planned, including studies into baseline noise levels and environmental effects. These tests will employ deployment-grade MF UFMs, with full cold optical stacks installed. At the time of this proceeding, the expectation is that the LATR will deployed to the field in early 2023, with first science observations occurring later that year.  \par
\section{Acknowledgements}
This work was funded by the Simons Foundation (Award \#457687, B.K.) and the University of Pennsylvania. 
Gabriele Coppi is supported by the European Research Council under the Marie Sk\l{}odowska Curie actions through the Individual European Fellowship No. 892174 PROTOCALC.

\bibliography{main} 
\bibliographystyle{spiebib} 

\end{document}